\documentclass[useAMS,usenatbib,usegraphicx]{mn2e}
\usepackage[totalwidth=480pt, totalheight=680pt]{geometry}

\title[Jet/disc coupling through energy reservoir]{Jet-disc coupling through a common energy reservoir in the black hole XTE~J1118+480}

\author[Malzac, Merloni \& Fabian]{Julien Malzac$^{1,2}$\thanks{E-mail:
malzac@ast.cam.ac.uk}, Andrea Merloni$^{3}$ and Andrew C. Fabian$^{1}$\\
$^{1}$Institute of Astronomy, Madingley Road, Cambridge, CB3 0HA \\
$^{2}$Centre d'Etude Spatiale des Rayonnements, CNRS-UPS, 9 Avenue du Colonel
Roche, 31028 Toulouse Cedex 4, France\\
$^{3}$Max-Planck-Institut f\"ur Astrophysik, Karl-Schwarzschild-Strasse 1, D-85741 Garching, Germany}

\begin{document}

\date{Accepted 1988 December 15. Received 1988 December 14; in original form 1988 October 11}
\pagerange{\pageref{firstpage}--\pageref{lastpage}} \pubyear{2002}

\maketitle

\label{firstpage}

\begin{abstract}

We interpret the rapid correlated  UV/optical/ X-ray
variability of XTE~J1118+480 as a signature of the coupling between the
X-ray corona and a jet emitting synchrotron radiation in the optical
band.
We propose a scenario in which the jet and the 
X-ray corona are fed by the same energy reservoir where
 large amounts of accretion power are stored
 before being channelled  into either the jet or 
the high energy radiation. This time dependent model reproduces
 the main features of the rapid multi-wavelength variability of
XTE~J1118+480.  
Assuming that the energy is stored in the form of magnetic field, we
find that the required values of the  model parameters 
are compatible with both a patchy corona atop a cold accretion disc and 
a hot thick inner disc geometry.
The range of variability timescales for the X-ray emitting plasma are consistent with the dynamical times of an
accretion flow between 10 and 100 Schwarzschild radii. On the other hand, 
the derived range of timescales associated with the
dissipation in the jet extends to timescales more than 10 times
larger, confirming the suggestion that the generation of a powerful
outflow requires large scale coherent poloidal field 
structures.
A strong requirement of the model is that the total 
jet power should be at least a few times
larger than the observed X-ray luminosity, implying a radiative
efficiency for the jet $\epsilon_{\rm j} \la 3 \times 10^{-3}$.
This would be consistent with the overall low radiative efficiency of the
source. We present independent arguments showing that the jet
probably dominates the energetic output of all accreting black
holes in the low-hard state.
\end{abstract}


\section{Introduction}
The high energy spectrum ($>$ 1 keV) of accreting stellar mass 
black holes in the low/hard state can be roughly described by a power-law
with photon index $\Gamma \sim 1.4-2$, and a nearly exponential
cut-off at a characteristic energy $E_{\rm c}$  of 
 a few hundred keV (see e.g. Tanaka \& Lewin 1995; Gierlinski et
 al. 1997; McClintock \& Remillard 2004).
Such a spectrum is generally interpreted as due to thermal
Comptonisation in a plasma with electron temperature
 $kT_{\rm e}\sim$ 100 keV
and Thomson optical depth $\tau\sim$ 1 (see e.g. Poutanen 1998). 
There are two possible explanations for the presence of this very hot
plasma (often called {\it corona}). It could be either a geometrically
thick, optically thin innermost part of the accretion flow
 (Shapiro, Lightman \& Eardley 1976; Narayan \& Yi 1994) 
or a collection of small scale active regions located atop a cold,
geometrically thin and optically thick accretion disc 
(Haardt, Maraschi \& Ghisellini 1994), 
possibly powered by magnetic reconnection.

When the X-ray luminosity increases above a few percent of the
Eddington luminosity ($L_{\rm Edd}$),
accreting black holes are observed to switch from the low/hard 
state to the so-called high/soft-state (see e.g. McClintock \&
Remillard 2004).
Then the X-ray luminosity is dominated by a strong thermal component
originating from a geometrically thin, optically thick disc.
Current observations show that the power-law component is steeper ($\Gamma \sim 2.5-3$) and much less
luminous than in the hard state, suggesting the
disappearance of the Comptonising plasma in this state.  

Recent multi-wavelength observations
of accreting  black holes in the hard state have shown the presence of
an ubiquitous flat-spectrum radio emission (see e.g Fender 2004), that may
extend up to infrared and optical wavelengths. 
The properties the radio emission indicate it is likely  produced
by synchrotron emission from relativistic electrons in compact,
self-absorbed jets (\citeauthor{bk79}, 1979; \citeauthor{hj88}, 1988). 
This idea was confirmed by the discovery 
of a continuous and steady milliarcsecond 
compact jet around Cygnus X-1  (\citeauthor{sti01}  2001).
Moreover, in hard state sources a tight 
correlation has been found between the hard X-ray and radio luminosities, holding over more than three decades in luminosity 
(Corbel et al. 2003; Gallo, Fender \& Pooley 2003).
In contrast, during high-soft state episodes the sources appear to be
radio weak (Tananbaum et al. 1972; Fender et al. 1999; \citeauthor{cor00} 2000), 
suggesting that the Comptonising medium of the low/hard 
state is closely linked to the continuous ejection of matter in 
the form of a small scale jet.

Merloni \& Fabian (2001) have pointed out that the energy content of
the electrons in the Comptonising medium is too low  to account
for the observed X-ray luminosities. The Comptonising electrons 
have to be tightly connected to an energy reservoir where large
amount of accretion power is stored before being transfered to them,
 and ultimately radiated. Independently,  the 
rapid X-ray variability in Cyg X-1 also suggests the presence of
of energy reservoirs (Negoro et al. 1995; Maccarone \& Coppi 2002).
Because angular momentum transport in
 accretion flow is most likely due to magneto-rotational instability
 (MRI, see Balbus \& Hawley, 1998), a natural candidate for the energy
 repository is the magnetic field amplified in the disc by the
 MRI-turbulent flow. However, if the dissipation of the tangled field 
(via magnetic reconnection) and non-adiabatic turbulent heating
preferentially energize the protons, rather than the electrons,
the hot Comptonising plasma could become two-temperature, and the
protons themselves act as the main energy reservoir
(\citeauthor{dbf97}, 1997). 

Models and simulations of jet production (Blandford \& Znajek 1977;
Blandford \& Payne 1982; Meier 2001)  indicate that jets are driven 
by the poloidal component of the magnetic field.
Therefore  storage of energy into  magnetic structures  
driving the jet and powering the Comptonising electrons 
is the most straightforward
explanation for the observed corona-jet association
\footnote{See however Markoff, Falcke \& Fender (2001) and
Georganopoulos, Aharonian \& Kirk (2002) for alternative
interpretations involving dominant X-ray emission from the jet.}.
In this context, the corona would constitute the location where 
the jet is launched.  
These idea were
developed by several authors  
in the context of geometrically thin and/or thick accretion flows 
(Meier 2001; Livio, Pringle \& King 2003) as well as accretion disc coronae
 (Merloni \& Fabian 2002).

Besides the radio/X-ray correlation  observed on long ($>$ 1 day)
timescales, there are  indications that the corona-jet 
coupling operates on timescales as short as a few seconds or less.  
The best example is provided by the X-ray nova XTE J1118+480
(Remillard et al. 2000; McClintock et al. 2001a).
During its outburst in 2000, this black hole showed all the X-ray
properties of hard state sources.     
Fast optical and UV photometry  has shown rapid optical/UV
flickering presenting
complex correlations with the X-ray variability (Kanbach et al. 2001; Hynes et
al. 2003, hereafter K01 and H03 respectively). 
This correlated variability cannot be caused by reprocessing 
of the X-rays in the external parts of the disc.
Indeed, the optical flickering occurs on average on shorter
time-scales than the X-ray one (K01), and reprocessing models fail to 
fit  the complicated shape of the X-ray/optical cross correlation 
function (H03).
Spectrally, the jet emission  seems to
 extend at least up to the optical band (McClintock et al. 2001b; Chaty
et al. 2003, hereafter C03), 
although the external parts of the disc may provide an
important  contribution to the observed flux at such
wavelengths.
The jet  activity is thus the most likely explanation for the rapid
observed optical flickering. For this reason, 
the properties of the optical/X-ray correlation 
in XTE J1118+480 might be of primary importance for the understanding 
of the jet-corona coupling and the ejection process.
 
The simultaneous optical/X-ray observations are described at length in
a number of papers (K01; Spruit \& Kanbach 2001; H03; Malzac et
al. 2003, hereafter M03). As discussed in these works, the observations are very challenging
 for any accretion model. The most puzzling pieces of evidence 
are the following:   
(a) The optical/X-ray Cross-Correlation Function (CCF) shows the optical band lagging the X-ray by $~$0.5
s, but with a dip 2-5 seconds in advance of the X-rays (K01); 
(b) The correlation between X-ray and optical light curves 
appears to have timescale-invariant properties: 
the X-ray/optical CCF maintains a similar, but rescaled, shape on
timescales ranging at least from 0.1 s to few 10 s (M03);
(c) The correlation does not appear to be  triggered by
 a single type of event (dip or flare) in the light curves; instead, as was
 shown by M03,  
 optical and X-ray  fluctuations of very different shapes, amplitudes
 and timescales are correlated in a similar way, such that 
the optical light curve is related to the time derivative of the X-ray
one.
Indeed, in the range of timescales where the coherence is maximum,
the optical/X-ray phase lag are close to $\pi/2$, indicating that the
two lightcurves are related trough a differential relation.
Namely, if the optical variability is representative 
of fluctuations in the jet power output  $P_{\rm j}$, 
the data suggest that the jet power scales roughly like $P_{\rm j} \propto
-\frac{dP_{\rm x}}{dt}$, where $P_{\rm x}$ is the X-ray power.

Here we will show that, if indeed there is a common energy reservoir feeding
both the jet and the corona,
 this differential relation is naturally satisfied, provided
that the jet power dominates over the X-ray luminosity.

We will first present the energy reservoir model
and suggest a simple physical scenario for the energy
reservoir and jet disc/coupling (Section \ref{sec:reserv}).
We  will then present a time dependent 
model that captures the main features
of the multi-wavelength variability observed in XTE J1118+480,
and discuss our main results obtained by comparing
the properties of the simulated lightcurves with the observations 
(Section \ref{sec:results}). 
Section \ref{sec:flow} will be devoted to a discussion of the
constraints on the nature of the accretion flow in XTE J1118+480
derived from both spectral and temporal analysis, while in Section
\ref{sec:jet}, we make an attempt to generalize these results to other
accreting black hole sources.
Finally, we summarize our conclusions in section \ref{sec:conc}.

\section{The energy reservoir model}
\label{sec:reserv}

\subsection{A simple analogue}

The time-dependent model that we have developed is complicated in
operation and behaviour. In order to gain some insight into its
operation consider a simple model consisting of a tall water tank with
an input pipe and two output pipes, one of which is much smaller than
the other. The larger output one has a tap on it. The flow in the
input pipe represents the power injected in the reservoir $P_{\rm i}$,
that in the small output pipe the X-ray power $P_{\rm x}$ and
in the large output pipe the jet power $P_{\rm j}$.

If the system is left alone the water level rises until the pressure
causes $P_{\rm i}=P_{\rm j}+P_{\rm x}$. 
Now consider what happens when the tap is opened
more, causing $P_{\rm j}$ to rise. The water level and pressure
(proportional to $E$) drop causing $P_{\rm x}$ to reduce. If the tap is then
partly closed, the water level rises, $P_{\rm j}$ decreases and $P_{\rm x}$
increases. The rate $P_{\rm x}$ depends upon the past history, or integral
of $P_{\rm j}$. Identifying the optical flux as a marker of $P_{\rm j}$ and the
X-ray flux as a marker of $P_{\rm x}$ we obtain the basic behaviour seen in
XTE\,J1118+480.

In the real situation we envisage that the variations in the tap are
stochastically controlled by a shot noise process. There are also
stochastically-controlled taps on the input and other output pipes as
well. The overall behaviour is therefore complex. The model shows
however that the observed complex behaviour of XTE\,J1118+480 can be
explained by a relatively simple basic model involving several energy
flows and an energy reservoir.

\subsection{A magnetic energy reservoir ?}
\label{sec:magres}
 This simple model is largely independent of the physical nature of the
energy reservoir. In a real accretion flow, the
reservoir could take the form of either electromagnetic energy stored in the
X-ray emitting region, or thermal (hot protons) or turbulent motions. The
material in the disc could also  constitute a reservoir of
gravitational or rotational energy behaving as described above.
 
In order to be more specific, let us outline the possibility of
jet disc coupling through  a \emph{magnetic} energy reservoir.
For the sake of simplicity, we assume that the main energy reservoir
 for the radiating electrons
is indeed the magnetic field. This would correspond to a system in
which the heating of the protons by the MHD turbulence is
negligible. Quataert (1998) and Quataert and Gruzinov (1999), have
shown that this is the case if magnetic pressure is not much smaller
than gas pressure (low plasma $\beta$ parameter), and we will
therefore assume this is indeed the case here.
Then, the power channelled into the particle heating (through magnetic
reconnection) and escaping
the corona as X-ray radiation can be written as 
\begin{equation}
\label{eq:px}
P_{\rm x}=(v_{\rm d}/R_{\rm x}) E,
\end{equation} 
where $v_{\rm d}$ is the field dissipation speed, which
depends on the details of the dissipation process, $R_{\rm x}$ is the
typical size of an X-ray emitting region and $E=(B^2/8\pi) V$ is the
total magnetic energy contained in the hot phase ($V$ is its volume).

The jet power is instead related to the {\it poloidal} field
component. For the purpose of simple estimates, the MHD jet power can
be written as (\cite{lop99}): 
\begin{equation}
\label{eq:pj}
P_{\rm j}=(B_{\rm p}^2/8\pi)A R_c\Omega,
\end{equation}
where $A$ is the area of the disc threaded by
the poloidal field and $R_c \Omega$ is the typical rotational velocity of
the field lines. If the field in the disc is amplified by MRI
turbulence and dissipates mainly in the hot coronal phase
 (i.e. it does not possess a large-scale external component)
, then its poloidal component can be expressed as $B_{\rm p}
\simeq h B$, where $h=H/R_c$ is the scaleheight of the hot phase
(Livio et al. 1999; Meier 2001; Merloni \& Fabian 2001). The
jet power can then be rewritten:
\begin{equation}
\label{eq:pj2}
P_{\rm j}= \frac{A}{V} h^2 R_c\Omega E.
\end{equation}

For the sake of simplicity, we assume that the jet power 
is taken mainly from the {\emph tangled} magnetic
field.  This implies that the outflow is accelerated by magnetic
dissipation processes and/or that the magnetic energy of the corona is
carried away by an essentially electromagnetic outflow. In fact, 
it is possible that the jet is powered directly by the disc 
rotational energy without significant field dissipation.
This would be the case whenever the poloidal field is coherent on
large enough scales to exert a significant torque on the disc (see e.g.
the model of King et al. 2003).
However, such a mechanism for angular momentum transport in the disc
competes with MRI, so that when large scale fields extract energy and angular
momentum from the disc, the MRI dynamo switches off, also draining  the
magnetic energy reservoir.

Summarizing, we will make the assumption that both the jet and X-ray 
power are tapped from the same magnetic energy
reservoir. Note that from  equations \ref{eq:px} and  \ref{eq:pj2} both jet
and X-ray powers scale \emph{linearly} with the total reservoir energy.
The total power extracted from the magnetic field in the hot phase is
then:
\begin{equation}
\label{eq:ptot}
P_{\rm x}+P_{\rm j}=\frac{B^2}{8\pi}\left(\frac{V v_{\rm d}}{R_{\rm x}}
  + A h^2 R_c\Omega\right). 
\end{equation}
If we define the relative fractional power of the jet $\eta=P_{\rm j}/(P_{\rm x}+P_{\rm j})$, we
then find that (see also \citeauthor{mf02}, 2002)
\begin{equation}
\label{eq:eta}
\eta=\left(1+\frac{V}{A R_{\rm x}}\frac{v_{\rm d}}{R_c\Omega}h^{-2} \right)^{-1}.
\end{equation}

The discussion so far makes no distinction between
a corona made of a collection of active phases atop a cold disc and a continuous
hot inner flow. Indeed the differences show up in the
geometrical factors in the above equations. In the case of a
structured corona, we consider $N$ cylindrical active region (magnetic
tubes) located atop the disc at distances ranging from $3R_{\rm S}$ to
$R_{\rm c}$.
The typical radius $R_a$ and height $H_a$ of the active regions scale
linearly with their the distance $R$ from the black hole, 
we note $a=R_a(R)/R$ and $h=H_a(R)/R$.
We assume that their radial distribution scales
like $1/R$ so that the covering factor of the corona is independent of
distance. Then the average radius of an active region is $R_{\rm
x}=aR_{\rm c}/\ln{(R_{\rm c}/3R_{\rm S})}$, $A=(\pi/2) N  a^2 R_{\rm
c}^2/\ln{(R_{\rm c}/3R_{\rm S})}$, and $V= (\pi/3) N a^2 h R_{\rm
c}^3/\ln{(R_{\rm c}/3R_{\rm S})}$. 

 On the other hand, in the thick disc
case, we have $R_{\rm x}=R_c$,  $A=\pi R_{\rm c}^2$ and $V=(2/3)\pi h
R_{\rm c}^3$, where here $R_{\rm c}$ is taken to coincide with 
the outer radius of the thick disc and a
constant $H/R$ (wedge-like geometry) is assumed.

\subsection{Time dependent model}
\label{sec:model}

In a stationary flow, the extracted power $P_{\rm j}+P_{\rm x}$ would be perfectly
balanced by the power injected into the magnetic field, which is, in
the most general case, given by 
the difference between the accretion power and the power
advected into the hole and/or stored in convective motions\footnote{
In fact, the secular evolution of Convection Dominated Accretion Flow
leads to accumulation of mass at the outer disc boundary, and thus to
a non stationary flow. However, if we limit ourselves to 
observations which are shorter than a typical
outburst duration, we can regard convection as an alternative escape route
for the graviational energy dissipated by the accreting matter out of the
region of interest (where X-ray and optical emission are produced).}: $P_i
\simeq \dot M c^2 - P_{\rm adv,conv}$. However, observations of
strong variability on short time scale clearly indicate that the heating and
cooling of the X-ray (and optical) emitting plasma are highly
transient phenomena, and the corona is unlikely to be in complete 
energy balance on short timescales.
We therefore introduce a time-dependent equation governing the
evolution of its total energy $E$:
\begin{equation}
\dot E= P_{\rm i} - P_{\rm j} - P_{\rm x},
\label{eq:enbal}
\end{equation}
and we assume that all the three terms on the right hand side are time dependent.

Furthermore, we assume that the optical light comes mainly from 
synchrotron emission in the inner part of the jet. A contribution 
from X-ray light reprocessed in the external part of the 
disc could be present in the optical emission. However, this component
is  likely to be weak (cf. K01; H03). 

The (time averaged) total observed optical flux is:
\begin{equation}
O_{pt} \propto P_{\rm j}+ f_{\rm r} P_{\rm x}
\label{eq:O}
\end{equation}
where the normalization factor $f_{\rm r}$ depends on the geometry of
the system and on the optical wavelength, with $f_{\rm r}$ increasing
towards shorter wavelengths\footnote{In the fully time dependent model
we will consider, additional effects on the properties of the optical
emission should be considered: First of all there is a time delay
$\Delta$ between matter ejection and dissipation in the form of
optical photons.  As the jet is relativistic and the optical light is
produced at short distance ($\la 1000 GM/c^2$) from the hole, the
delay should be $\Delta \la 0.1$ s. Similarly, reprocessing will also
introduce a time delay of order of $0.1$ s and any variability of
$P_{\rm x}$ on time scales shorter than a few seconds will not be
apparent in the reprocessed light}.

We introduce the instantaneous dissipation rates $K_{\rm j}$ and $K_{\rm x}$~:
\begin{equation}
 P_{\rm j}(t)= K_{\rm j}(t)E(t)  
\end{equation}
\begin{equation}
 P_{\rm x}(t)= K_{\rm x}(t)E(t), 
\label{def:kx}
\end{equation}
and we consistently  denote $\langle K_{\rm j} \rangle$, $\langle K_{\rm x}
\rangle$, $\langle P_{\rm i} \rangle$, and  $\langle E \rangle$ 
as the time averaged values.
We define the average dissipation time:
\begin{equation}
\label{eq:tdiss}
T_{\rm dis}=(\langle K_{\rm j} \rangle+\langle K_{\rm x} \rangle)^{-1},
\end{equation}
such that if the energy reservoir is not fed (i.e. $P_{\rm i}=0$) its level
decays with an e-folding time $T_{\rm dis}$. Also, it is convenient to define 
\begin{equation}
\label{eq:fx}
f_{\rm x}=1-\eta= \langle K_{\rm x} \rangle T_{\rm dis},
\end{equation}
which is the average fraction of the total power that goes into the X-ray
emission.
In the framework of the magnetic reservoir of section \ref{sec:magres}, the average values of the dissipation rates 
are related by eqs. (\ref{eq:px}) and
(\ref{eq:pj}) to the (time time averaged) values of the 
physical parameter of the system. Namely: $\langle K_{\rm j}
\rangle=(Ah^2/V)R_{\rm c}\Omega$ and
$\langle K_{\rm x} \rangle=v_{\rm d}/R_{\rm x}$.

Obviously, the detailed physical modeling of the time evolution of
such a jet corona system is an  extremely complex problem that is
beyond the present computer capabilities.
Instead we will adopt a more phenomenological approach.
We will model the variability of the source
by assuming random fluctuations of $K_{\rm j}$, $K_{\rm x}$ and $P_{\rm i}$,
then compare the results with the observations and constrain the
properties of these fluctuations.

In general, both the instantaneous injected power $P_{\rm i}$ and the
dissipation rates $K_{\rm j}$ and $K_{\rm x}$ may depend on the amount of
energy $E$ stored in the reservoir. However, the fact that $P_{\rm i}$ and
$K_{\rm x}$ are required to vary on time scales different from those of
$P_{\rm x}$ (see below), suggest that they are rather independent of $E$.
Moreover, observations show that in black hole binaries and Seyfert
galaxies the X-ray amplitude of variability is linearly related to the
X-ray flux level (Uttley \& Mc Hardy 2001). In general, if $P_{\rm i}$,
$K_{\rm x}$ and $K_{\rm j}$ are either strongly dependent on $E$, or
correlated with each other, this would introduce a non-linear relation
between RMS amplitude and observed flux. The observations thus suggest
that the values of $P_{\rm i}$ and the dissipation rates $K_{\rm j}$ and
$K_{\rm x}$ at a given time are nearly independent of the level of the
reservoir energy (although $E$ does depend on the history of $K_{\rm
j}$ and $K_{\rm x}$). For these reasons, as well as for the sake of
simplicity, we will then assume that $K_{\rm j}$, $K_{\rm x}$ and $P_{\rm i}$
fluctuate randomly and independently, driving the fluctuations of the
energy reservoir.

For the specific form of the fluctuation, we use exponentially rising shot profiles:
\begin{equation}
s(t)=A\frac{\exp(t/\tau)-1}{\exp(1)-1}  \qquad {\rm for} \quad  t < \tau;
\end{equation}
This profile was chosen for its simplicity. Our results regarding the optical/X-ray correlation are not
sensitive to the shape of the individual flares. On the other hand
they will strongly depend on the amplitude, time-scales and
occurence rate of the shots.
The amplitude $A$  and the occurrence rate $\lambda$ of the random
shots are taken constant. 
These quantities are related to the average dissipation rate.
The shot duration $\tau$ is distributed within $\tau_{min}$ and
$\tau_{max}$ with a power law distribution 
$\rho(\tau) \propto \tau^{-p}$. 
These parameters constrain the fractional amplitude of variability and the 
shape of the power spectrum of the fluctuations that are imposed on the system.
The fractional RMS scales like $1/\sqrt{\lambda}$, and the power spectrum is a
power law ranging from frequencies $1/\tau_{max}$ to $1/\tau_{min}$
with a slope $\alpha=3-p$ (see e.g. Poutanen and Fabian 1999).
For $K_{\rm x}$, $K_{\rm j}$ and $P_{\rm i}$, the  parameters $\lambda$, $p$,
$\tau_{min}$, $\tau_{max}$ are, in general, not identical.
In presenting our results, the subscripts $x$, $j$ and $i$ will refer
to $K_{\rm x}$, $K_{\rm j}$ and $P_{\rm i}$ respectively.

For a specific set of parameters we first generate time series for
$K_{\rm x}$, $K_{\rm j}$ and $P_{\rm i}$, 
solve the time evolution of the
energy reservoir $E$ and then use it to derive the
 the resulting optical and X-ray light curves.
The ejected material travels from the corona to the shock region
or  photo-sphere where the optical photons are produced. As discussed above,
this introduces a time delay of $\Delta < 0.1$ s in the optical emission.
To model this delay, we simply shift the optical light curve 
by $\Delta$.
We then compute the X-ray and optical Power Density Spectra (PDS), Auto
Correlation Functions (ACF), their Cross-Correlation Function (CCF),
coherence and phase lag spectrum for comparison with the observed ones, in order
to obtain the largest possible number of different observational tests
for
our variability model.

\subsection{Main observational constraints}
\label{sec:constraints}
Before we proceed and illustrate the results of our simulations, a
remark is in place with respect to the requirement  any specific
realization of our model will have to fulfill in order to reproduce the
main observational characteristics. This will guide us in
the exploration of the vast parameter space of the model, as well as
provide us with useful insight on the physical interpretation of the
results.

We first note that 
combining equations (\ref{eq:enbal}) and  (\ref{def:kx}) we obtain for
the total instantaneous jet power the
following relation: 
\begin{equation}
P_{\rm j}=P_{\rm i} - (1 + \frac{\dot K_{\rm x}}{K_{\rm x}^2})P_{\rm x} -
\dot P_{\rm x}/K_{\rm x}.
\label{eq:jdom}
\end{equation}

We can see from this equation that the differential
scaling $P_{\rm j} \propto - \dot P_{\rm x}$, observed in
XTEJ1118+480, will be rigorously reproduced  
provided that:
\begin{itemize} 
\item $K_{\rm x}$ is a constant;
\item $P_{\rm i} - P_{\rm x}$ is a constant.\footnote{This
constant can differ from 0 because
 the differential scaling is observationally
demonstrated only for the \emph{varying}
fraction of the optical and X-ray fluxes.}     
\end{itemize}

It is physically unlikely that those conditions will be  exactly
verified. In particular, $P_{\rm x}$ is observed to have a large RMS
amplitude of variability of about 30 percent. 
However, the observed differential relation holds only roughly
and only for fluctuations within a relatively narrow range of
time-scales $1-10 s$. Therefore, the above conditions need only to be
fulfilled approximatively and for low frequency fluctuations ($> 1$s). 
In practice, the following requirements will be enough to make sure
that  the low frequency fluctuations of the right hand side  
of equation $\ref{eq:jdom}$ are dominated by $\dot P_{\rm x}$:
\begin{itemize} 
\item $P_{\rm x} \ll P_{\rm i}$, implying that the jet power, on average, dominates over the X-ray luminosity;
\item the amplitude of variability of $K_{\rm x}$ and $P_{\rm i}$ in the
$1$-$10$ s range is low compared to that of $P_{\rm j}$. In other words the
$1$-$10$ s fluctuations of the system are mainly driven by the jet activity.  
\end{itemize}

The first condition is crucial, and will be discussed in more detail
 in Section \ref{sec:flow}.
The second condition requires that the mechanisms for energy reservoir
filling  and  dissipation 
in the corona and  in the jet are occurring on quite different
timescales, and this will also provide us with additional constraints
on the global dynamical properties of the system.

\section{Results}
\label{sec:results}

In figure \ref{fig:simu184}, we show the results of a simulation with
$T_{\rm dis}=0.5$ s and $f_{\rm x}=0.1$ (see Table~\ref{tab:parameters} for the value of the other parameters).
The model produces an X-ray power spectrum with a plateau
up to $\sim$ 0.1 Hz and a power-law 
component with slope $\sim 1.4$ above that frequency, with 
most of the X-ray variability
occurring around 0.1 Hz. The optical PDS power-law has a flatter slope ($\sim 1$)
up to 1 Hz and then softens to a slope similar to
that of the X-ray PDS.
The resulting optical ACF is  significantly narrower than the
X-ray one. The full-width-at-half-maximum (FWHM) of the two ACFs
 differs by a factor $>2$. 
The overall coherence is low ($<0.4$): 
reaching a maximum in the 0.1--1 Hz range and decreasing rapidly
 both at lower and higher frequency. 
The  phase-lags 
are close to $\pi/2$ in the  0.1--1 Hz range and increase from 0 at
low frequencies up to $\pi$ at around 6 Hz. At higher frequencies the
phase lags spectrum is characterized by large oscillations.
 Finally the resulting CCF rises very quickly at positive optical lags,
peaks around $0.5$~s  (this is the post-peak)
and then declines slowly at larger lags.
The two bands appear to be anti-correlated at negative optical
 lags indicating a systematic optical dip 1-2 s before the X-rays
reach their maximum (pre-dip).

All these characteristics are observed in XTE J1118+480.
Moreover, the integrated X-ray RMS is 25 percent as observed. The optical
variability  is 20 percent slightly larger than the 16 percent reported by
M03 but this could be reduced if the expected 
 constant disc component was added.

Obviously the model parameter values were carefully
selected in order to reproduce these characteristics.
Given the complexity of the model, a full investigation of the
parameter space is premature. However,
it may be useful to illustrate the main effects of the different parameters 
 by considering several simple situations.

\subsection{Variability dominated by the jet dissipation rate $K_{\rm j}$}

\subsubsection{Jet dominated models  ($f_{\rm x}$$<<$1)}
We first  consider the case where the X-ray emission is energetically
negligible.
The jet fully drives the variability of the system in the limit of large 
$\lambda_{\rm i}$ and $\lambda_{\rm x}$, and low $f_{\rm x}$. In this limit, both
conditions for the applicability of the model discussed in the previous
section are exactly verified. We thus expect the scaling $P_{\rm j}
\propto -\dot P_{\rm x}$ to be realised. Indeed, we obtain phase lags that 
are close to $\pi/2$ independent of frequency.
In this limit, the resulting CCF is purely anti-symetric with pre-dip
and post-peak of identical amplitude and lag (contrary to what is observed). 
The optical lag of the post-peak is controled by $T_{\rm dis}$, which
sets
the time scale on which the energy reservoir responds 
to the imposed fluctuations.
At frequencies larger than $1/T_{\rm dis}$ the X-ray variability decreases 
because there cannot be any variability of the reservoir energy $E$ 
on time scales larger than $T_{\rm dis}$.
 The coherence function also decreases at frequencies above $1/T_{\rm dis}$.
On the other hand, at frequencies lower than $1/T_{\rm dis}$ the optical variability 
decreases because at such low frequencies the system evolves
in a quasi-static way. Since $P_{\rm i}$ is constant and $P_{\rm x}$ is
energetically negligible, the variability of $K_{\rm j}$ is compensated by the
variability of $E$ so that  $P_{\rm j}$ remains almost constant, too.

\subsubsection{X-ray dominated models  ($f_{\rm x}$$\sim$1)}

If $f_{\rm x}$ is large but the amplitude of variability of $K_{\rm x}$
and $P_{\rm i}$  are kept negligible by increasing $\lambda_{\rm x}$ and
$\lambda_{\rm i}$, then the power output can be dominated by the X-rays
but the variability is still driven by the jet. 

In this case, we still have $\pi/2$ phase-lags but only at frequencies $>1/T_{\rm dis}$.
 The effects of a large X-ray dissipation is the
appearance of phase-lags $~\pi$ at frequencies $<1/T_{\rm dis}$ indicating an
anti-correlation. This anti-correlation is also apparent in the CCF,
which tends to be negative and symmetric around zero lag.
The anti-correlation at low frequencies is due to the fact that on long
time-scales the system is always in equilibrium, but contrary to the
jet dominated case, X-ray losses are not negligible in this case. 
The reservoir energy $E$ changes in order to keep
the \emph{total} output constant (and equal to $P_{\rm i}$), but $P_{\rm j}$
does change.
 For this reason the fluctuations of the jet output are 
strictly anti-correlated to that of the X-ray
power. Such an anti-correlation is in clear conflict with the data,
the observed phase-lags indicating rather a correlation at low frequencies.
Thus, even if the jet drives the variability as in the previous
 example, we definitely need
$f_{\rm x}\ll1$ in order to reproduce the data. In practice, exploring a
 large volume of the parameter space we could not find 
a reasonable agreement with the data for $f_{\rm x}$ larger than  $\sim
0.2$. This is a very robust result of our time-dependent modeling, and
 we will examine its consequences in Section \ref{sec:flow}.

\subsection{Variability dominated by the coronal dissipation rate
 $K_{\rm x}$ and/or by the power input $P_{\rm i}$}

The situation where the variability is dominated by the X-ray 
dissipation rate in the corona is perfectly symmetric to the case
where the variability is dominated by the jet, studied above, 
with however an inversion of the sign of phase lags and time-axis in
the CCF plot. 

On the other hand,  fluctuations
of $P_{\rm i}$ introduce 
perfectly positively correlated fluctuations of $P_{\rm j}$ and $P_{\rm x}$.
If the fluctuations of $P_{\rm i}$ dominate we obtain 
X-ray and optical light curves that are perfectly correlated on all time scales with
identical power spectrum and unity coherence, in contrast with the observations.  

In general, when in a given range of timescales more than one component
(i.e. $P_{\rm i}$, $K_{\rm j}$ or $K_{\rm x}$) drives the variability, the main
effect is a strong reduction of the coherence function in the
corresponding range of frequencies.

\subsection{A realistic case}

Now that we understand the basic effects of the different parameters,
we get back to our 'realistic' model and explain how we were guided
toward this solution and how it reproduces  qualitatively all the
timing features observed in XTE J1118+480. 

As discussed above, reproducing the data requires 
the jet to dominate the energy budget and be the main driver of the
variability  
(at least for the 1--10 s fluctuations).
Therefore we set $f_{\rm x}$=0.1 and the parameters  $\lambda_{j}$, $p_{j}$, $\tau_{min\,j}$ and $\tau_{max\,j}$ were
 adjusted in order to have large enough 1-10 s fluctuations of the jet power.
These parameters also control the shape of the optical power spectrum.
In particular, the slope of the shot distribution  $p_{j}$ is chosen
so that it leads to the observed slope of the optical power spectrum
above 1 Hz.
 The dissipation time is set to $T_{\rm dis}= 0.5$  in order to match the observed
optical  lag in the CCF reported by K01.

The values of  $\lambda_{i}$, $p_i$, $\tau_{min\,i}$ and $\tau_{max\,i}$
were fixed so that the variability of $P_{\rm i}$ introduces some
correlated variability at low frequencies, enabling us to reproduce the
asymmetry of the CCF and the small  phase-lags observed at low 
frequency. The loss of coherence below 10 Hz is due to the
simultaneous large variability of $P_{\rm i}$ and $K_{\rm j}$ at low frequencies. 

Finally, we also  introduce fluctuations of the X-ray dissipation
rate at high frequencies. 
This enables us to reproduce the loss of coherence observed above one
Hz. Also  $\lambda_{x}$, $p_{x}$, $\tau_{min\,x}$ and $\tau_{max\,x}$
 were chosen in order to reproduce the X-ray power
spectrum at high frequencies which is otherwise too steep.

A propagation lag of $\Delta=$0.05 s then provides the observed high frequency behaviour
of the phase lags. If the propagation lag is  neglected, the phase lag
spectrum is flat at high frequencies with phase lag at a constant
value $\sim \pi/2$.   
The constant time delay introduces an additional phase lag $\psi=2\pi f \Delta$. As
the time-scale of the fluctuations approaches that of the time delay 
this additional phase lag becomes dominant. The overall phase lag thus
increases and reaches $\pi$ at $\sim 6$ Hz. The phase lags are defined
only in the range [$-\pi$,$\pi$] and can be only measured modulo
2$\pi$. Thus, at higher frequencies the phase lag shifts to -$\pi$ then
increases linearly to $\pi$ and so on, producing large 'oscillations'
in the pahse-lag  spectrum. A similar  behaviour of the phase lags above 1
Hz is observed in XTE J1118+480 (see M03, Fig.~5),
indeed suggesting a propagation lag of $\Delta\sim 0.05$ s.
On the other hand, the overall effects of propagation delays 
 on the CCF and coherence function are negligible.

\subsection{The effects of reprocessing}\label{sec:reprocessing}

We now consider the effects of a possible reprocessing component
 from the disc. We model this component by convolving the X-ray
 lightcurve with a transfer function $T_r$ describing the time delay and
variability smearing due to reprocessing.
We use the following transfer function~:
\begin{eqnarray}
T_r(t) &\propto &
\frac{t}{\delta}\exp(-t)\exp\left[\left(\frac{t}{\delta}\right)^{0.01}-\frac{t}{\delta}\right]
 \nonumber\\
&+&0.28\left[1-\exp\left(-10\frac{t}{\delta}\right)\right]\exp\left[-(t/5)^{20}\right]
\end{eqnarray}
where $t$ is measured in seconds and the
reprocessing time delay $\delta=0.1$ s.

This function is a rough analytic approximation to the theoretical
transfer functions shown in H03.

\begin{figure*}
\includegraphics{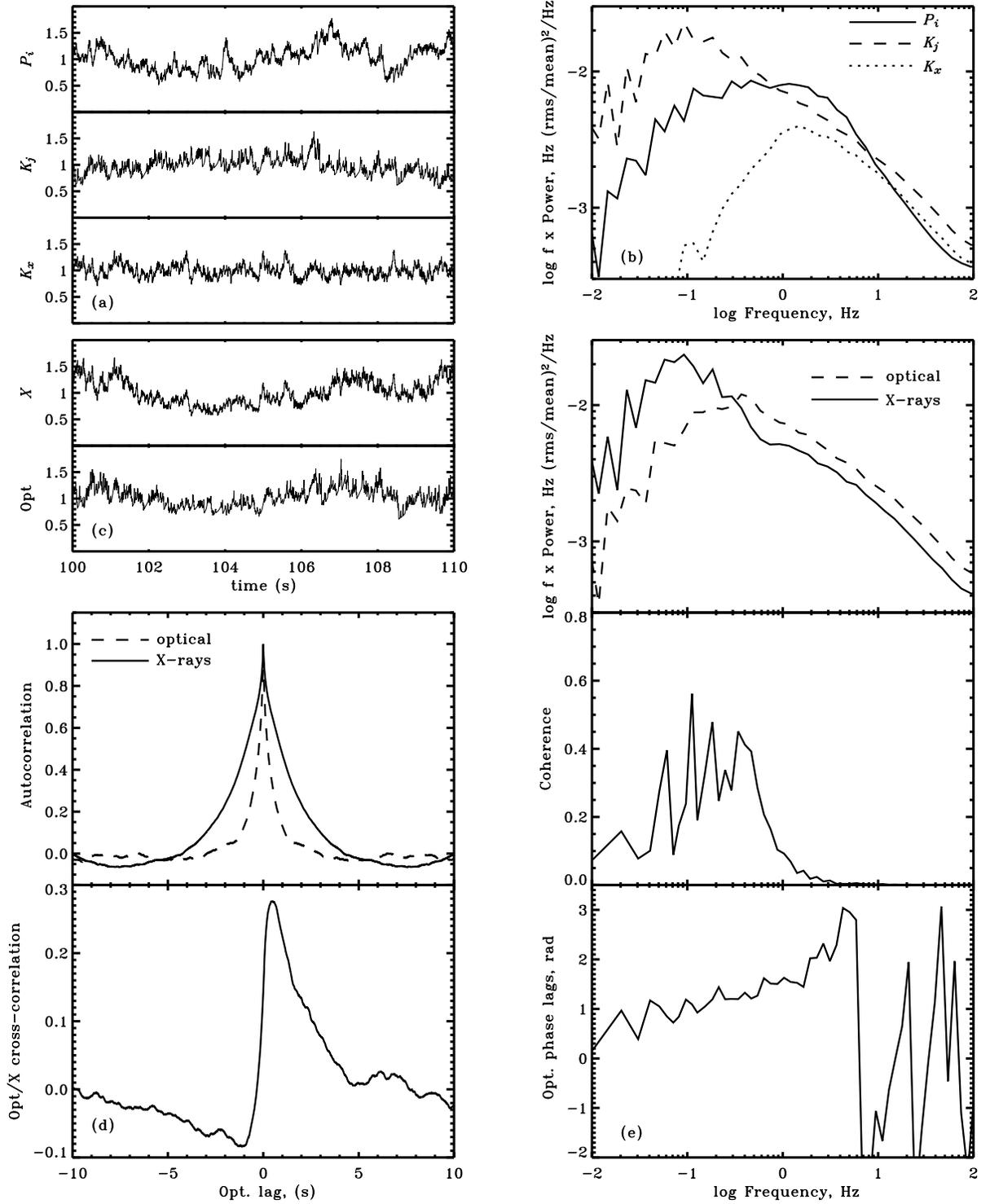}
\caption{Results of simulations for parameters given in
table~\ref{tab:parameters}. Sample time series (panel a) and power
spectra (panel b) of $P_{\rm i}$,
$K_{\rm j}$, $K_{\rm x}$, resulting X-ray and optical fluxes light curves
(panel c),  X-ray/optical autocorrelation and cross-correlation
functions (panel d), power spectra, coherence and phase-lags (panel e).} 
\label{fig:simu184}
\end{figure*}

\begin{figure*}
\includegraphics{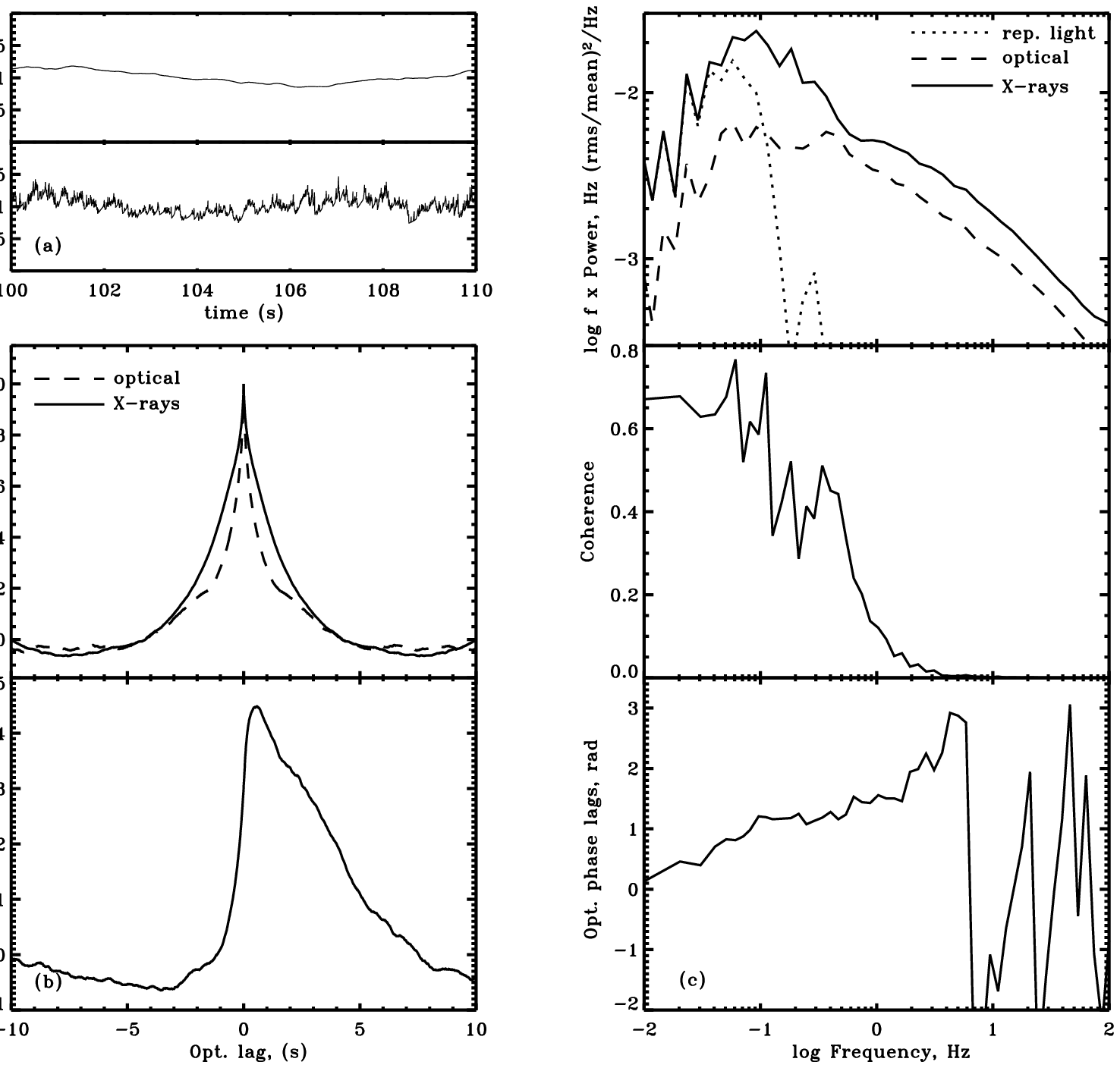}
\caption{Same simulation as in Fig.~\ref{fig:simu184}
 but adding a disc reprocessing component
 accounting for 30~percent of the optical flux on average (see section~\ref{sec:reprocessing}).
Panel a: sample light curves of the reprocessed component and total optical
Panel b: X-ray/optical autocorrelation and cross-correlation
functions.
Panel c:  power spectra of the reprocessed component, total optical
and X-rays (top), X-ray/optical coherence and phase-lags (bottom).} 
\label{fig:simu185}
\end{figure*}

The reprocessed component is added to the optical light curve.
Its normalization is parametrized by the ratio of the average flux
of the reprocessed component to that of the jet component.
In the simulation shown in Figure~\ref{fig:simu185}, we take this
ratio to be 0.5, keeping all other parameters' values identical to
those of the
previous simulation. The main effect is to add a low frequency
component to the optical lightcurves highly correlated to the
X-ray one. By comparison with Figure~\ref{fig:simu184}, the optical ACF
is broader, the coherence is higher at low frequency
and the optical pre-dip tends to disappear, the optical lags are
shorter in particular at low frequency. 

The increase of the reprocessed component is associated with
an increase in the coherence function, and the X-ray and optical ACF
and PDS becoming similar and shorter optical lags.
We expect to observe this evolution if we look at shorter and shorter
 wavelengths where the disc emission increasingly dominates the total
optical/UV flux. Indeed H03 report similar trends in the
dependence of the correlation upon wavelength.
We also note that at low
frequencies ($<0.1$ Hz), H03 observe a large coherence  in
the far UV domain, while M03 report almost zero coherence. 
This could be interpreted as the effects of reprocessing in the far
UV that would become negligible at optical wavelengths.
\begin{table}
\begin{tabular}{ccc}
\hline
$T_{\rm dis}=0.5$ s       &  $f_{\rm x}=0.1$         &   \\
$\lambda_{\rm i}=100$ s$^{-1}$     & $\lambda_{\rm j}=50$ s$^{-1}$     & $\lambda_{\rm x}=1000$ s$^{-1}$\\
$p_{\rm i}=2.1$           & $p_{\rm j}=1.4$          & $p_{\rm x}=1.1$  \\  
$\tau_{min\,i}=0.1$ s & $\tau_{min\,j}=0.01$ s & $\tau_{min\,x}=0.01$ s\\
$\tau_{max\,i}=7$ s   &$\tau_{max\,j}=10$ s   &  $\tau_{max\,x}=0.5$
s\\
\hline
RMS $P_{\rm i}=0.20$        &   $K_{\rm j}=0.27$   &      $K_{\rm x}=0.11$  \\

RMS $E=0.23$          &   $\mathrm{Opt}=0.21$     &      $X=0.26$\\
\hline
\end{tabular}
\caption{ The model parameters used in the simulations of
Fig.~\ref{fig:simu184}. The two last rows show
the resulting fractional amplitude of the dissipation rates, energy
reservoir, optical and X-ray fluxes. In model \ref{fig:simu185}
some reprocessing was added and  the optical fractional RMS is then reduced to 0.16.}
\label{tab:parameters}
\end{table}

\section{The nature of the low-luminosity accretion flow in XTE J1118+480}
\label{sec:flow}
\subsection{Constraints from the time variability study}
The detailed analysis of the complicated structure of the
multi-wavelength variability of the source have allowed us to select a
region of parameter space for which our model successfully
reproduces all the main observational pieces of evidence. The
parameters are summarized in Table~\ref{tab:parameters}. 
In principle these parameters can contrain the dynamic and
geometry of the accretion flow. However this would require a detailed 
physical model for the jet/disc coupling. Here we will assume that, as
suggested in section~\ref{sec:magres}, the
magnetic field constitutes the main energy reservoir and investigate the
consequences for the accretion flow. 

Let us start by examining the two parameters that reflect the time-averaged
global energetics of the system: $T_{\rm dis}$ and $f_{\rm x}$.
From Eqs. (\ref{eq:px}), (\ref{eq:pj2}), (\ref{eq:tdiss}) and
(\ref{eq:fx}), we see that they are related to the physical parameters
of the system:

\begin{equation}
\frac{v_{\rm d}}{R_{\rm x}}=\frac{f_{\rm x}}{T_{\rm dis}}
\label{eq:vd}
\end{equation}

\begin{equation}
\frac{A}{V}h^2 v_{\rm K}(R_{\rm c})=\frac{1-f_{\rm x}}{T_{\rm dis}},
\label{eq:geom}
\end{equation}
where we have assumed that the disc rotation law is Keplerian and we
 denote $v_{\rm K}(R_{\rm c})$ the Keplerian speed at the radius
$R_{\rm c}$.

In the estimates below, we will assume, as we do throughout the paper,
a black hole mass of $10 M_{\odot}$ for XTE J1118+480.
Equation (\ref{eq:vd}) is a constraint on the magnetic field dissipation
speed in the hot phase: for $f_{\rm x}=0.1$ and $T_{\rm dis}=0.5$ it can be
 rewritten as:
\begin{equation}
\label{eq:vdis}
\frac{v_{\rm d}}{c}\simeq 2 \times 10^{-5} r_{\rm x},
\end{equation}
where $r_{\rm x}=R_{\rm x}/R_{\rm S}$. 

On the other hand, Equation (\ref{eq:geom})  
constrains the corona/hot flow  scaleheight,
which should be defined  by selecting one of the two possibility for
the hot phase geometry. It turns out that for both geometries we
obtain the same relation:
\begin{equation}
h \simeq 1.7 \times 10^{-1} \frac{0.5}{T_{\rm dis}}\frac{\eta}{0.9}
\left(\frac{r_{\rm c}}{100}\right)^{3/2}
\label{eq:lxm}
\end{equation}
where $r_{\rm c}=R_{\rm c}/R_{\rm S}$.

In the case of the thick inner flow, self-consistency
would require a substantial scaleheight of the hot flow. If, for
example, we require $h\ga 0.1$, we obtain a constraint on the radial
extent of the inner hot flow $r_{\rm c} \ga 70$, which, incidentally, appear
to be consistent with the spectroscopically-inferred inner radius of the truncated cold disc (Esin et
al. 2001; Chaty et al. 2003).

The same spectroscopic constraints also apply to the patchy corona case
(see next section) and the radial extension of the corona above the disc
should also be at least $r_{\rm c}\sim 100$. Then for typical inferred values
of the size of the active regions ($r_{\rm x} \sim$ a few, see e.g Haardt,
Maraschi \& Ghisellini 1994; Di Matteo, Celotti \& Fabian 1999), the
aspect ratio of the cylindrical regions writes:

\begin{equation}
\frac{h}{a} \simeq 1.6 \; \frac{0.5}{T_{\rm dis}}\frac{\eta}{0.9}
\frac{3}{r_{\rm x}}\left(\frac{r_{\rm c}}{100}\right)^{5/2} \frac{\ln(r_{\rm c}/3)}{\ln(100/3)}.
\label{eq:hsa}
\end{equation}
It appears to be slightly larger than unity, as required
to reproduce the hard X-ray spectrum (see e.g. Malzac, Beloborodov 
\& Poutanen 2001). Note that the aspect ratio is very sensitive to the
coronal radius. If  $r_{\rm c}$ is as large as 350, as inferred by Chaty et
al. (2003) we obtain a quite large ratio $h/a\simeq 27$, then 
internally generated cyclo-synchrotron radiation should dominate over
reprocessing as a source of seed photons for the Comptonisation process.

Moreover, combining the expression that relates the observed X-ray
luminosity (\ref{eq:px}) with the constraints obtained above on the
field dissipation speed, we can estimate the value of the magnetic field.
In the case of the thick inner flow we obtain:
\begin{eqnarray}
 B  \simeq 3.2 \times 10^{6}
\left[\frac{f_{\rm x}}{0.1}\frac{\eta}{0.9}\right]^{-1/2}\frac{T_{\rm
dis}}{0.5 {\rm s}}\left(\frac{r_{\rm c}}{100}\right)^{-9/4} \;\hbox{G}
\end{eqnarray}

From the lower limit on the inner disc radius derived above, we conclude that
$B \la 7 \times 10^6$ G. In the case of the patchy corona:
\begin{eqnarray}
 B  \simeq 2.5 \times 10^{7}
\left[\frac{f_{\rm x}}{0.1}\frac{\eta}{0.9}\frac{\ln(r_{\rm c}/3)}{\ln(100/3)}\frac{N}{10}\right]^{-1/2}
\nonumber \\ 
\frac{T_{\rm dis}}{0.5 {\rm s}} \frac{3}{r_{\rm x}} \left(\frac{r_{\rm c}}{100}\right)^{-5/4} \;\hbox{G}
\end{eqnarray}

We note that the
RMS variability amplitude should scale as 
$N^{-1/2}$, so that the magnetic field 
intensity is directly proportional to the observed variability level.

Finally, let us discuss the different variability timescales of the
dissipation rates, as fixed in our fiducial model. On the one hand, 
the range of timescales on which  the variability of $K_{\rm x}$ is maximal
are between 50 and 1 Hz, corresponding to the Keplerian frequencies
between 10 and 100 $R_S$ (for a 10 $M_{\odot}$ black hole). This is
consistent with the X-rays being produced in the hot phase of the 
accretion flow (either in a geometrically thick disc or in the
corona). In general the dissipation of the tangled magnetic field into
particle heating is
essentially a local process acting on timescales comparable to 
the local dynamical time.
Therefore the minimum and maximum variability
timescales of $K_{\rm x}$ reflect the smallest and largest radii where we have
substantial energy in the hot phase/corona.

On the other hand, we see from Figure~\ref{fig:simu184} that the
variability of $K_{\rm j}$ is high on much wider range of timescales than $K_{\rm x}$,
 from 0.01 to 10 seconds. As discussed in sections \ref{sec:constraints} and
\ref{sec:results}, the
data require that at low frequencies
the fluctuations of $K_{\rm j}$ dominate 
  over the other sources of variability. 
The physical interpretation is that generating a powerful 
outflow requires large-scale coherent poloidal field 
structures threading a significant fraction of the inner disc
 surface and extending also in the vertical direction.
Such structures may take more time to build and destroy.
 So, locally, the timescale
of variation of $K_{\rm j}$ can go from the dynamical time itself to 
larger timescales (in the rare cases where the field
builds up its poloidal component to a larger scale).
Based on simple arguments, Livio, Pringle \& King (2003)  
estimate that the time-scale for establishing a change 
in the scale of the poloidal component of the magnetic field
through dynamo processes is larger than the dynamical time 
scale by a factor $t_{jet}/t_{dyn} \sim 2^{1/h}$.
Since our model indicates  $t_{jet}/t_{dyn} \sim 10$,  this would
imply $h\sim 0.3$, which appears to be in agreement with the
independent estimate provided by equation \ref{eq:lxm}.

We also note that the variability of $K_{j}$  induces
large amplitude fluctuations of the energy reservoir that in turn lead to
an important X-ray variability on time-scales much larger than the
dynamical time. 
In fact, most of the observed X-ray variability is caused
by the jet activity,  providing an explanation for the long observed
time-scales. Recently King et al. (2003) have proposed a detailed
physical model for the X-ray variability of accreting black holes
 that considers a similar situation where large-scale ejection events 
modulate the strong X-ray emission from the inner disc. 

\subsection{Constraints from time averaged spectroscopy}

We have shown in section \ref{sec:results}
that there is an important condition for our
a model to work, namely that 
the jet power should dominate over the X-ray luminosity.
We will now discuss how realistic this assumption is.

First, let us examine the observed energetic output of
XTE~J1118+480. We stress that this is the best source available for
such kind of study, given its unsurpassed spectral coverage. 
 In the following we will express all
the luminosities in units of the Eddington luminosity ($L_{\rm
  Edd}=1.3\times 10^{38} (M/M_{\sun})$ erg s$^{-1}$, with $M=10 M_{\odot}$).
The numbers given below are taken from the results of the
multi-wavelength spectral analysis of Chaty et al. (2003).
These authors decomposed the overall SED into three components:
\begin{itemize}
\item The {\it jet component} has a total observed radiative luminosity of $L_{\rm j}=
  3.8 \times 10^{-5}$, assuming isotropy and integrating the power-law
  emission from the radio to the optical band. 
In the optical/UV the emission should  start to be dominated by the cold
accretion disc, but the possibility that the jet spectrum extends up into the
UV domain cannot be ruled out. Indeed, the fact that H03 observe similar
correlated variability in the UV and optical lend support to this hypothesis. 
If this is the case, the estimated jet luminosity 
 can increase by up to a factor of 8. 
\item For the standard {\it cold accretion disc} we have: $L_{d}=1.7 \times 10^{-3}$.
This estimate was obtained
by Chaty et al. (2003) with a fit of the optical to EUV spectrum with a multicolour
blackbody model. The inner disc temperature was found to be
very low $kT\sim 20$ eV, suggesting that either the inner  disc is truncated
at $\sim 350$ $R_{\rm S}$, or that the bulk of the accretion power is driven
away in a coronal outflow and the disc is left very cold.
\item Finally, for the luminosity of the X-ray emitting plasma, or
  {\it corona}, we have: $L_{\rm x}=1.2 \times 10^{-3}$, assuming isotropy.
\end{itemize}

Summing up all the three above spectral components, we find that the   
bolometric luminosity of the source is:
\begin{equation}
\label{eq:bol}
L_{\rm bol}=L_{\rm j}+L_{\rm d}+L_{\rm x} \simeq 3 \times 10^{-3}
\end{equation}
   
The total mass accretion rate onto the black hole
can be estimated from the observed luminosity of the cold disc
component. We define the Eddington-scaled accretion rate 
$\dot m =\epsilon \dot M c^2/L_{\rm Edd}$ where $\epsilon$ is the
Newtonian  accretion efficiency $\epsilon=1/12$.
If the cold, geometrically thin and optically thick disc is indeed 
truncated at $R_{\rm in}=350$~$R_{\rm S}$, as the spectral analysis of
Chaty et al. (2003) suggests, then standard accretion formulae (Frank,
King and Raine 2002) give:
\begin{equation}
\label{eq:rin_mdot}
\dot m= 4 \epsilon \frac{R_{\rm in}}{R_{\rm S}} L_{d}\sim 0.2. 
\end{equation}
On the other hand, if the accretion disc extends down to the innermost
stable orbit ($R_{\rm in}=3 R_{\rm S}$), but is sandwiched by a powerful corona, where a fraction
$f_{\rm H}$ of the power is dissipated, then we can estimate that
\begin{equation}
f_{\rm H}=\left(1+\frac{L_{d}(1-\eta)}{L_{\rm x}}\right)^{-1}.
\end{equation}
The $\eta$ parameter is only poorly constrained,
but in  any case $f_{\rm H}$ should be very
large in order to explain the low inner disc temperature (Merloni, Di Matteo \&
Fabian 2000). Taking $\eta\simeq 0.9$ as indicated by the time dependent model, we obtain
$f_{\rm H}=0.88$, and, for the total mass accretion rate,
\begin{equation}
\dot m= 4 \epsilon \frac{R_{\rm in}}{R_{\rm S}} \frac{L_d}{(1-f_{\rm H})} \sim 0.015.
\label{eq:adcmdot}
\end{equation}
In both cases, the estimated accretion rate is much larger than the observed
bolometric luminosity. 
Thus,  most of the accretion power \emph{is not radiated} and
the source  appears to be \emph{radiatively inefficient}.

The important issue, however, would be to determine whether the missing accretion
power escapes the system in the (low radiative efficiency) jet or in
other forms of non-radiative losses, such as a slow wind, or large
scale convective motions, or advection into the black hole. 
The answer to this question resides in the exact determination of the
jet kinetic power.
Unfortunately, there are major uncertainties in  
this determination, mainly because the jet radiative efficiency
is not known.
The jet is expected to be a poor radiator because most of the energy
is lost in adiabatic expansion, thus,  
although the radiation from the jet represents a small fraction of the
bolometric luminosity the jet could dominate the energetics.
Both theory and observations
indicate that the efficiency can be extremely
low (as low as $\epsilon_{\rm j} \sim 10^{-4}$ see Celotti and Fabian
1991), but all present  estimates are  model dependent.
In general they indicate radiative efficiencies of the order of 
 $\epsilon_{\rm j}\sim 0.01$. For the case of  XTE~J1118+480 this
would already imply that the total jet power dominates over the X-ray
luminosity. Obviously, if the jet efficiency
is lower, or the jet spectrum extends to shorter wavelength, or beaming
effects are important, the jet dominance can easily be much larger.
As discussed above, the analysis of our time dependent modeling
strongly requires $f_{\rm x}\la 0.1$, corresponding to 
a jet efficiency $\epsilon_{\rm j} \la 3\times 10^{-3}$.

In the case of the truncated disc plus 
inner hot inner flow model, the derived $\dot m$ sets an
upper limit to the jet power, in the case where all the
accretion power is lost in the jet:  $P_{\rm j}/L_{\rm x}\sim \dot m/ L_{\rm x} \sim 200$.
On the other
hand, if the jet power is comparatively 
modest e.g. $P_{\rm j}/L_{\rm x}\sim 10$ as assumed in the simulation of
Fig.~\ref{fig:simu184}, this would imply that, although the jet dominates over the
X-ray emission, it does not dominate as a power sink. If 
the dominant power sink is, for example,  advection onto the black hole,  the mass
accretion rate we derive appears somewhat large to be consistent with
ADAF models (Narayan \& Yi 1994). 
However, it is important to  notice that the observed 
accretion rate is very sensitive to the value of the
inner disc radius (see eq. \ref{eq:rin_mdot}), which is actually poorly
constrained. 
Moreover, the
critical accretion rate for ADAF models above which the ADAF solution
breaks down scales as $\alpha^2$ (Rees et al. 1982; Narayan \& Yi
1995), and the unknown standard $\alpha$
parameter could be larger than usually considered leading to
the existence of solutions at higher $\dot m$ than previously thought.
 Finally, we note that the dynamics of the ADAF solutions coupled with jets
have not been worked out yet. The structure of the accretion flow as
well as  the (in)stability of the solution at high
$\dot m$ are likely to  be strongly affected by the presence of the jet.
Although an advection-dominated accretion flow is in many aspects
consistent with our analysis, it should certainly differ considerably 
from the standard solution. 

In the case of the accretion disc plus patchy corona (or coronal
outflow) model the $\dot m$ derived assuming $\eta=0.9$ 
is consistent with \emph{all} of the
non-radiated power driving the jet. However, when we attempted to fit
the optical-UV spectrum of XTEJ1118+480 with the disc corona solution
of Merloni \& Fabian (2002), we found that it was not possible to
produce inner disc temperatures as low as observed ($\sim$ 20 eV).
The coronal outflow model could be reconciled with the EUV spectral
data only if 
an even  larger fraction of the power was extracted from the disc into the
corona/jet, requiring $f_{\rm H}$ to be very close to 1. 
However,  
in the framework of magnetic coronae powered by MHD turbulent disc,
having $f\sim 1$ also requires a very  large 
(magnetic) viscosity (Merloni 2003), and implies an even stronger jet 
dominance of the total energy budget of the source.

An alternative way to obtain a low inner disc temperature together
with reasonable value of $f_{\rm H}$ would be that the cold disc does
not extend down to the last stable orbit. Instead, below a few 
tens of $R_{\rm S}$ the disc is truncated and the flow becomes very
inefficient. However, such a picture would be very similar to the
truncated disc plus hot inner flow scenario discussed
above, except that the jet would be launched from the external
accretion disc corona.

\section{Some wider implications of the model}
\label{sec:jet}
\subsection{Comparison with other sources}
The behaviour of XTE\,J1118+480 demonstrates that the accretion flow
onto a black hole can lead to a strong jet with little disk emission.
A similar situation occurs in the massive elliptical galaxy M87 where
powerful jets are associated with an otherwise weak galactic nucleus.
The Bondi accretion rate onto the supermassive black hole in M87,
deduced from the properties of the surrounding gas with Chandra
observations (Di Matteo et al 2002), can give the observed jet power,
deduced from the cavities made by them in the surrounding intracluster
medium, provided that accretion efficiency is 0.1 or more. Assuming
that the flow is in a steady state, such a high efficiency argues that
most of the accreting gas flows to within a few gravitational radii of
the black hole.  Since thick hot flows are unstable to mass loss over
a wide range af radii
( Begelman \& Blandford 1999; Quataert \& 
Gruzinov 2000; Stone, Pringle Begelman 2000),
 this favours a magnetically-dominated cold disk flow in the
case of M87. By similarity it argues for a magnetically-dominated cold
disc in XTE\,J1118+480 as well. The reservoir and presumably the base
of the jet must also lie within a few gravitational radii of the black
hole in both cases.

As mentioned in the Introduction, the low state of many Galactic Black
Hole systems appears to contain a powerful jet. When systems drop
below about one per cent of the Eddington accretion rate the disc may
be magnetically dominated (Merloni \& Fabian 2002; Livio, Pringle \&
King 2003), with a fast jet taking most of the gravitational energy
released. This may also apply to massive black holes, as illustrated
above with M87. The issue of how apparently well-fed massive black
holes in galactic nuclei are feeble sources of electromagnetic
radiation has been a puzzle for some years (Fabian \& Canizares 1988;
Fabian \& Rees 1995; Di Matteo et al 2001, 2003; Loewenstein et al
2001; Pellegrini et al 2003). Most of these objects do however have
jetted radio emission (Franceschini, Vercellone \& Fabian 1998) and a
solution in which the accretion power is principally carried away by
jets is a strong possibility. Unless the energy contained in black
hole spin is a key factor, then magnetically-dominated discs should be
part of that solution.

Further time-dependent studies of the multiwavelength behaviour in
other stellar-mass and massive black hole systems may show similar
behaviour to that of XTE\,1118+480. Some examples which already offer
tantalizing behaviour are the Galactic microquasar GRS1915+105
(Mirabel et al., 1998; Klein-Wolt et al., 2002), the Galactic binary GX339-4 which has also displayed
rapid optical flickering (Motch et al 1982) and the blazar 3C120 which
shows X-ray dips before radio jet events are seen (Marscher et al
2002).

\subsection{Jet-dominated sources in the low/hard state}

Whatever the actual structure of the accretion flow, we have shown
that during the outburst of XTE~J1118+480, 
the total kinetic jet power should dominate over the X-ray
luminosity, and could possibly be the dominant repository of the
accretion power.
There are additional independent arguments in favour of jet dominance 
in low/hard state sources and in XTE~J1118+480 in particular.
Based on the observed radio flux ($L_{\rm R}$) and X-ray correlation observed
in hard states sources (Falcke \& Biermann 1996; Gallo, Fender \& Pooley, 2003),
as well as on standard synchrotron formulae (Heinz \& Sunyaev, 2003), 
Fender, Gallo \& Jonker (2003) have shown that, provided that
advection into the black hole horizon and/or convective motions do not
store a large fraction of the accretion power, there should exist a
critical accretion rate, $\dot m_{\rm cr}$, below which 
an accreting black hole is jet-dominated.

The exact value for the critical accretion rate could be inferred from the
observations, if we knew the total jet power at a
certain X-ray luminosity, and is given by $\dot m_{\rm cr}=2 P_{\rm
  j}^2/L_{\rm x}$, corresponding to a critical X-ray luminosity
$L_{\rm x,cr}=\dot m_{\rm cr}/2$. 
Fender et al. (2001) derived a lower limit for the jet to X-ray power ratio 
in XTE~J1118+480: $P_j/L_{x}=0.2$, and 
FGJ03 used this conservative estimates to determine the value of the
critical rate $\dot m_{\rm cr} \simeq 7 \times 10^{-5}$.
However such a low value of the critical luminosity leads to several
problems.

First,  as shown in FGJ03, during the transition 
from a disc to a jet-dominated state, the dependence of the X-ray luminosity
 on the accretion rate
changes from being $L_{\rm x} \propto \dot m^2$, 
the right scaling for \emph{radiatively
inefficient flows}, to $L_{\rm x} \propto \dot m$, the scaling for
\emph{radiatively efficient flows} (see Fig.~1 of FGJ03).
This would imply that with $L_{\rm x}\sim 10^{-3}$,  XTE~J1118+480
 should be a  \emph{radiatively efficient} system. As discussed above, 
there is however strong observational evidence of the contrary.

Second, during the transition from jet to disc dominated state, 
the jet power changes from $P_j \propto \dot m$ 
which is the natural scaling expected in most jet models, to the
seemingly unphysical $P_j \propto \dot m^{1/2}$.

Furthermore, black holes in the hard state should show some kind of
spectral transition in the X-ray band at the critical luminosity 
$L_{\rm x,cr} \sim 3 \times 10^{-5}$, due to the drastic changes in
emission mechanisms that are needed to account for the different
scalings of $L_{\rm x}$ with the accretion rate. The observations of
low/hard state sources at such low luminosities are few and hard to
perform, however no indication of any dramatic spectral change in any
hard state source down to quiescent level has ever been reported (Kong
et al., 2002; Hameury et al., 2003)

In fact, the only physical transition that we do actually observe is the
transition between the hard and the soft state that occurs at luminosities
of at least a few percent of Eddington luminosity (Maccarone 2003).
We believe that, if the above mentioned difficulties are to be solved,
then $\dot m_{\rm cr}$ has to
correspond to luminosities that are comparable to, or larger than,
hard-to-soft state transition luminosities.

For the case of XTE~J1118+480, instead of using the lower limit 
for the jet to X-ray power ($P_j/L_{x}=0.2$), we can adopt 
the much larger value $P_j/L_{x}\sim 10$
required by our variability model. Then 
we find $\dot m_{\rm cr} \sim 0.2$, involving a transition at
$L_{\rm x,cr}=\dot m_{\rm cr}/2 \sim 0.1$.
 This is in agreement with the idea that the transition from jet
dominated to X-ray dominated states occurs at luminosities similar or
slightly higher than the hard to soft state transition.
We stress that the arguments of FGJ03  holds only as long as the
$L_{\rm R}-L_{\rm x}$ correlation is valid i.e. not in the soft-state.
There is thus no reason to associate the disc dominated sate of FGJ03
to the soft state. 

If, as we suggest, the critical luminosity $L_{\rm x,cr}$ is
at or above the hard to soft state transition luminosity, then the disc
dominated state of FGJ03 does not exist.
Moreover, although one can speculate that the transition to soft state 
may be be triggered by the
loss of the jet dominance, there is a priori no reason for the critical
luminosity $L_{\rm x,cr}$ to be coincident with the hard to soft
state transition luminosity. 
In other words, the observed presence of a hard-to-soft state transition 
does not provide an upper limit to the jet power, but only a
lower limit.

We also note that the above discussion was made neglecting any
contribution from
advection and other possible non-radiative losses.
However, in jet-dominated sources the jet power scales
proportionally to the accretion rate, as is the case for any adiabatic
loss. Therefore the effects of advection (and/or convection) 
would not change the scalings of jet and X-ray
powers provided that we replace the accretion rate $\dot m$ 
with $\dot m_{\rm eff}=\dot m (1-f_{\rm ad})$,
 where $f_{\rm ad}$ is the fraction of accretion power that is advected
into the black hole. In particular, the critical X-ray luminosity
$L_{\rm x,cr}$,  although corresponding to a larger
'real' $\dot m$, would remain unchanged. 

To conclude, the central requirement of our model i.e. that in
XTE~J1118+480 the jet power dominates over 
the X-ray luminosity, appears to be supported
by the observations of hard state black holes. 
Then, if the arguments of FGJ03 are correct, 
an important consequence of the jet dominance in XTE~J1118+480
is that \emph{all hard state sources are jet-dominated} (in the sense
that the jet power dominates over the X-ray power). 
This jet dominance also implies that 
\emph{all hard state sources should be radiatively inefficient}.

It is not clear whether most of the accretion
energy is channeled into the jet or across the black hole event horizon. 
The $L_{\rm x}\propto \dot m_{eff}^2$ dependence of the X-ray
luminosity (also observed in low-luminosity AGN, as inferred from the
multivariate radio-X-rays-mass correlation, Merloni, Heinz \& Di
Matteo 2003; Falcke, K\"ording \& Markoff 2003) 
is similar to what is predicted by radiatively inefficient
accretion model.  
The reason for this inefficiency could be advection into the jet 
as well as advection into the black hole.

\section{Conclusions}
\label{sec:conc}

We have shown that the puzzling optical/X-ray correlations of
XTE~J1118+480,
 can be understood in terms of a common energy reservoir for
both the jet and the Comptonizing electrons.
For illustration purpose, we have presented a specific shot noise
variability model that reproduces {\it all} 
the main observed features of the multi-wavelength correlated variability.

Our time dependent model is
fairly general, and our main conclusions hold regardless of the
specific geometrical and dynamical properties of the system.
The main results can be summarized as follows:
\begin{itemize}
\item{Any energy reservoir model for XTE J1118+480
requires that the total jet power dominates
 over the X-ray luminosity. In particular, assuming that the compact 
jet synchrotron emission extends up to the optical band, this implies
a radiative efficiency for the jet $\epsilon_{\rm j}\la 3\times
10^{-3}$. Following the same line of arguments as FGJ03, we showed that
this situation is likely and probably represents a common feature of all
black holes in the low-hard state}
\item{The range of typical variability timescales of dissipation rate
of the X-ray  emitting plasma is consistent with the dynamical times of an
    accretion disc between $\sim$10 and 100 Schwarzschild radii (for a
    10 $M_{\odot}$ black hole). As expected, the X-ray variability can
    be associated with either a hot, thick inner accretion flow or
    with a patchy corona.}
\item{As jet launching  requires large-scale coherent
magnetic structures, the energy dissipation in the jet should vary on
longer time scales. Indeed, the derived range of timescales associated with the
dissipation in the jet extends to timescales more than 10 times larger
    than the X-ray emitting one.} 
\item{By combining the information obtained from our time variability
    model with (time averaged) spectral analysis we conclude that,
    whatever the accretion geometry, the whole disc-corona-jet system
    must be radiatively inefficient. It is therefore possible, in
    principle, that the system is energetically dominated by the jet,
    as suggested by FGJ03, although 
whether the bulk of the accretion power is advected into the black
hole or into the jet remains at present an open question. }
\item{In terms of specific dynamical models, we conclude that the
    observed properties of XTE J1118+480 during its low/hard state
    outburst are consistent with either an inner hot, quasi spherical,
radiatively inefficient flow, from which the jet originates (Meier 2001),
surrounded by a geometrically-thin, optically-thick cold disc, or with
a powerful patchy, outflowing corona on top of an extremely cold
standard thin disc. In the first case, multicolour disc fits
to the UV/EUV spectrum indicate a very large inner disc radius,
implying a large total accretion rate ($\dot m\simeq 0.2$), which might
be in conflict with the hypothesis of standard advection dominated
flow theory. In the second case, in order to reproduce the very low
inner disc temperature, an (uncomfortably) 
extreme coronal power is needed, together
with substantial relativistic bulk motion of the coronal plasma, both
possibly associated with very high magnetic viscosity.}
\end{itemize}

\section*{Acknowledgements}
JM acknowledges financial support from PPARC.
We are grateful to Annalisa Celotti and Jim Pringle for useful discussions.

\end{document}